\documentclass{report}
\usepackage{svcon2e}

\def\gapx{\lower 2pt
\hbox{$\buildrel>\over{\scriptstyle{\sim}}$}}
\def\lapx{\lower 2pt
\hbox{$\buildrel<\over{\scriptstyle{\sim}}$}}

\def\chapteraddress{Department of Physics, University of Toronto,\\Toronto,
Ontario, Canada M5S 1A7}

\begin{document}
\pagenumbering{arabic}
\chapter{BEC AND THE NEW WORLD OF COHERENT MATTER WAVES}
\chapterauthors{Allan Griffin}
\noindent\chapteraddress

\vskip 12pt

After decades of effort to produce an atomic Bose condensate, this was 
finally achieved in June, 1995 in a laser-cooled magnetically trapped gas 
of $^{87}$Rb atoms by Eric Cornell and Carl Wieman at JILA (University of 
Colorado and N.I.S.T.).  As of July, 1999 (4 years later), there are now 
over 20 experimental groups around the world who can routinely produce and 
study such atomic Bose condensates.  In addition, over 1000 theoretical 
papers have been published!

I will give three {\it introductory} lectures on this new {Quantum Phase 
of Matter}, which I think will continue to be a growth point of 
fundamental physics research for the next decade and will also be the 
source of new technologies based on using this source of {\it coherent 
matter waves}.
A brief sketch of my three lectures is as follows:
\begin{enumerate}
\item A BRIEF HISTORY OF BEC STUDIES.
\begin{itemize}
\item Before 1995, going back to pioneering work of Einstein (1925) and 
Fritz London (1938).
\item Introduce the concept of a {\it Bose macroscopic wavefunction} 
$\Phi({\bf r}, t)$ describing a Bose condensate and the key 
Gross-Pitaevskii equation of motion for $\Phi({\bf r}, t).$   
\item Review some of the results obtained over the last 4 years - and 
indicate why trapped Bose gases are so interesting, possibly even more 
than superfluid $^4$He and BCS superconductors, which also involve Bose 
condensation.
\end{itemize}
\item DYNAMICS OF A PURE CONDENSATE $(T\ll T_{BEC})$.
\begin{itemize}
\item Time-dependent Gross-Pitaevskii (1961) equation of motion.
\item Crucial effect of weak interatomic interactions.
\item Collective oscillations of the condensate.
\item The Bose condensate as a {\it classical} quantum object!
\end{itemize}
\item DYNAMICS OF COUPLED CONDENSATE AND NON-CONDEN-SATE COMPONENTS 
(TWO-FLUID HYDRODYNAMICS):
\begin{itemize}
\item Derivation of a {\it quantum} Boltzmann equation for the 
non-conden-sate excited atoms and a generalized GP equation for the 
condensate.
\item More complex behaviour than in superfluid $^4$He.
\item Comparison with the well-known Landau two-fluid theory (1941).
\end{itemize}
\end{enumerate}
Further references which are relevant for these lectures are:
\begin{enumerate}
\item BEC Homepage (maintained by a BEC theorist, Mark Edwards):
\newline
http://amo.phy.gasou.edu/bec.html/.  This widely-used homepage has played 
a crucial role in BEC research.
\item A recent article \cite{Dalgiorpit} by Dalfovo, Giorgini, Pitaevskii 
and Stringari is an authoritative review of recent theory on atomic Bose 
condensates and expands on the material I cover in Sections 
\ref{sec:Overview} and \ref{sec:Dynamics}.
\item  Articles in a book \cite{Ingstrwie}, {\bf Bose-Einstein  
condensation in atomic gases},  {\it Proceedings of the International 
School of Physics 
``Enrico Fermi''}, ed by M. Inguscio, S. Stringari and C. Wieman.  This 
contains many review articles on current BEC research.  In particular, I 
call attention to (all these can be downloaded from the LANL website under 
cond-mat):
\begin{itemize}
\item W. Ketterle, D.S. Durfee and D.M. Stamper-Kurn, ``Making, probing 
and understanding Bose-Einstein condensates'' - a 100 page review of 
recent experiments \cite{ketterle}.
\item A. Griffin ``A brief history of our understanding of BEC: From Bose 
to Beliaev'' \cite{Agrif}.
\item A. Griffin, ``Theory of excitations of the condensate and 
non-condensate at finite temperature'' \cite{Agrif}.
\item A.L. Fetter, ``Theory of a dilute low-temperature trapped Bose 
condensate.''  A very detailed analysis at $T = 0$.
\end{itemize}
\item A long article by Zaremba, Nikuni and Griffin \cite{Zarnikgrif}  on 
the non-equilibrium dynamics of trapped Bose gases at finite temperatures. 
This expands on the material covered in Section~\ref{sec:Coupled} of these 
lectures.
\end{enumerate}

\section{AN OVERVIEW OF PAST AND RECENT WORK}
\label{sec:Overview}

\subsection{Some history before 1980}

Einstein predicted that a non-interacting gas of atoms (Bosons) would 
undergo a {\it phase transition} at low temperatures, when a {\it 
macroscopic} $(0(N))$ number of atoms occupy the lowest energy level (in a 
uniform ideal Bose gas, this is the the zero momentum single-particle 
state ).  His work was inspired by a novel derivation of the Planck 
distribution for photons by Bose in 1924.  The basic physics of this phase 
transition is worked out in every text in statistical mechanics 
\cite{Khuang}.  The simplest way of estimating $T_{BEC}$ is to note that 
the transition occurs when

\begin{equation}
\lambda_T\, \gapx \ d = {\rm average \ distance\ between\ atoms} \sim {1 
\over n^{1/3}}, \label{eqoverview1}
\end{equation}
where $\lambda_T$ is the thermal De Broglie wavelength of a gas of atoms 
at temperature $T$,

\begin{equation}
\lambda_T \equiv \left({2\pi\hbar^2 \over mk_BT}\right)^{1\over 
2}.\label{eqoverview2}
\end{equation}
The criterion in (\ref{eqoverview1}) is equivalent to 
$n\lambda_T^3\,\gapx\, 1$, while a more careful analysis \cite{Khuang} 
gives $n\lambda^3_T = \zeta(3/2) = 2.612.$  One sees that $\lambda_T 
\rightarrow $ large as $T\rightarrow 0.$  When $\lambda_T\, \gapx \ d,\, $ 
all the atoms become {\it correlated} and the gas exhibits new collective 
behaviour (even in {\it absence} of interactions).
Using (\ref{eqoverview1}),  one finds that
\begin{equation}
k_B T_{BEC} \sim {2\pi\hbar^2\over m}n^{2/3}.\label{eqoverview3}
\end{equation}
Below $T_{BEC}$, the number of atoms in the ${\bf p} = 0$ single-particle 
state increases and is given by the well-known formula
\begin{equation}
{N_c(T)\over N} = \left[1 - \left({T\over 
T_{BEC}}\right)^{3/2}\right].\label{eqoverview4}
\end{equation}
At $T=0$, all the atoms in an ideal gas are in this ${\bf p}=0$ state 
(this state is the Bose condensate in non-interacting 3D gas).

Nothing much happened until 1938.  Then the neglected work of Einstein was 
re-discovered and developed by Fritz London \cite{Flondon}, who suggested 
that it might be the basis of an explanation for the strange effects 
noticed in liquid $^4$He at $T_c\sim 2.17K.$  London's suggestion was 
based on the fact that the $^4$He atom was a ``composite'' Boson $(S = 0)$ 
and the formula in (\ref{eqoverview3}) gives  $T_{BEC}\sim 3K$ if we used 
the density for liquid $^4$He.
L. Tisza used London's idea and suggested (somehow!) that the condensate 
atoms act in a coherent way - a {\it new collective degree of freedom 
moving without friction}.  This ``picture'' led to a rudimentary two-fluid 
model that could explain experiments showing superfluidity (especially by 
Kapitza as well as by Allen and Meisner) as a counterflow of the 
superfluid and the normal fluid.  In a dilute, weakly interacting Bose 
gas, these two components are the condensate and non-condensate, 
respectively.

Until the 1960's,  the theory of BEC in interacting systems was dominated 
by efforts to use it to understand superfluid $^4$He.  The London-Tisza 
scenario was essentially correct, but to formulate it properly needed 
field-theoretic many body techniques and the concept of broken-symmetry, 
which were only developed in the period 1957-1965.  In this period, a 
large amount of work was done on a {\it toy} problem, a dilute weakly 
interacting Bose gas, since a Bose liquid like superfluid $^4$He was too 
difficult to deal with theoretically.  These early studies \cite{Grif2} 
are the foundation of our current understanding of trapped atomic gases.

A phenomenological theory of superfluid $^4$He was introduced by Landau in 
1941, based on the idea of quasiparticles (phonon-roton spectrum) and a 
two-fluid superfluid hydrodynamics \cite{Land}.  This brilliant theory has 
been very successful and is the basis of modern descriptions of superfluid 
$^4$He \cite{Khalatnikov}.  However,  it made no explicit mention of BEC 
or even the fact that $^4$He atoms obeyed Bose statistics.  Only in the 
1960's did it become clear that Landau formulation had it's {\it 
microscopic} basis in the existence of a condensate macroscopic 
wavefunction,
\begin{equation}
\Phi({\bf r}, t) = <{\hat\psi}({\bf r})> = \sqrt{N_c({\bf r}, t)} 
e^{i\theta({\bf r}, t)},\label{eqoverview5}
\end{equation}
where $\hat\psi({\bf r})$ is the quantum field operator (see 
Section~\ref{sec:Dynamics}).   
This concept was first formally introduced by Beliaev in 1957 \cite{Bel2}, 
extending the pioneering work of Bogoliubov in 1947 \cite{Nnbog}.  The 
superfluid motion is associated with the gradient of the phase of this 
two-component order parameter,
\begin{equation}
e^{i\theta} = e^{i(\theta_0 + {\bf 
r}\cdot\mbox{\boldmath$\nabla$}\theta)}\ ; \ {\bf k}_s \equiv 
{\mbox{\boldmath$\nabla$}}\theta\equiv{m{\bf v}_s\over\hbar} 
\label{eqoverview6}
\end{equation}
The essential relation between superfluidity and $\Phi({\bf r}, t)$ is 
simply and elegantly described in the classic monograph by Nozi\`eres and 
Pines \cite{Gorkov}.
For further discussion of the development of our current understanding of 
Bose condensates, see the review article by Griffin in Ref. 
\cite{Ingstrwie}.

It might be useful to make a brief digression here on the BCS theory of 
superconductors based on formation of Cooper pairs (total spin $S=0),$ 
which was developed in 1957.  In the BCS theory, superconductors exhibit 
the same kind of {\it macroscopic} quantum behaviour as superfluid $^4$He, 
as first argued in the late 1930's by F.London.  In the Gor'kov version of 
the  BCS theory, the spin-singlet Cooper pair order parameter
\begin{equation}
\Phi({\bf r}, t)=<{\hat\psi}_\uparrow({\bf r}){\hat\psi}_\downarrow({\bf 
r})>
\label{eqoverview7}\end{equation}
is the equivalent of 
\begin{equation}
\Phi({\bf r}, t) = <{\hat\psi({\bf r})}>
\label{eqoverview8}\end{equation}
in Bose superfluids.  
The essential equivalence of these two systems was obscured by the 
complexity of the original many-particle BCS-wavefunction, due to the 
large spatial size of overlapping Cooper pairs.  As a result, Cooper pairs 
{\it form} and {\it condense} into a coherent state at {\it same} 
temperature.  It was only in the 1980's that theorists (Leggett, 
Nozi\`{e}res and others \cite{Randiera}) realized that for small Cooper 
pairs (tightly bound and hence high $T_c$), the BCS theory smoothly goes 
over to theory of a weakly interacting Bose gas of non-overlapping Cooper 
pairs.
Recently a gas of $^{40}$K atoms (a composite fermion) have been 
laser-cooled at JILA to slightly below the Fermi temperature 
\cite{Demarco}.  There are ways to make the interaction attractive (see 
later) and then one might look for the formation of Cooper pairs 
\cite{HTCStoof} of ultra-cold Fermi atoms!

\subsection{More recent developments (1980-1995)}
Since the 1970's, there has been increasing interest by experimentalists 
to find a ``pure'' form of BEC, namely, in a low temperature gas.  The two 
early candidates for Bosons were excitons (electron-hole pairs) in 
semiconductors and spin-polarized hydrogen atoms (see articles by Greytak 
and by Wolfe et al. in the book mentioned in Ref. \cite{Randiera}). A gas 
of $H_\uparrow$ atoms was predicted to be stable as a gas even at $T=0$.  
This is because the atoms cannot combine since there is no bound state of 
the interatomic potential between two spin-polarized H atoms.  Thus, one 
cannot form liquid or solid phase.  Many people got interested in BEC in 
$H_\uparrow$ gas, including theorists \cite{TJGreytak}.  

Several  of the key ideas that led to success in alkali atoms in 1995 grew 
out of the pioneering work on $H_\uparrow$ gas in the 1980's.
However 3-body interactions become increasingly important at higher 
densities and these cause spin flips, allowing formation of H-molecules. 
High densities were needed since cooling was by cryogenic methods, which 
could reach $\sim 10^{-4}$K but no lower.  BEC in $H_\uparrow$ gas was 
finally produced at MIT in June, 1998, after almost 20 years of work 
\cite{Physicstoday}.  Unfortunately, the alkali atom condensates are much 
easier to create and study, and appear to be more interesting gases.

Since the early 1990's, attention has focussed on the alkali atoms: Li, 
Na, K, 
Rb, Cs.  The atoms are Bosons, with an even number of neutrons.  The 
strategy was to use laser-cooling to get to very low temperatures, where 
the low density gas would Bose-condense.  The essential idea behind laser 
cooling is that when an atom absorbs a photon, it slows down.  In the 
summer of 1995, BEC was announced by three groups led by:

\begin{itemize}
\item C. Wieman and E. Cornell (JILA), using $^{87}$Rb atoms 
\cite{Andensmatwiecor}.
\item W. Ketterle (MIT), using $^{23}$Na atoms \cite{Davmewand}.
\item R. Hulet (Rice University), using $^7$Li atoms \cite{Bradsach}.
\end{itemize}

\noindent Parenthetically, it is now the general feeling  that the 
original Rice data, interesting as it was,  did not give an unambiguous 
signature of a (very small, since the interaction is attractive) 
condensate in $^7$Li gas \cite{Bradsachhul}.

Alkali atoms are perfect for BEC studies.  They have a magnetic moment, 
and hence can be trapped by magnetic fields.  They essentially have a 
``one-electron'' structure.  They are thus simple atoms, and have been 
well studied by atomic physicists.  One can easily selectively flip the 
``spin'' of higher energy trapped atoms.  These ``hot'' atoms are then 
quickly ejected from the magnetic trap and the remaining atoms quickly 
thermalize to a lower temperature.  This ``evaporative cooling'' is very 
efficient and quickly brings one into the temperature region required for 
BEC.

It is useful to mention a few experimental facts about the magnetic traps 
currently in use.  As it turns out, these traps are well described as a 
harmonic potential
\begin{eqnarray}
V_{ex}({\bf r})&=&{1\over 2}m\omega^2_0 r^2 \ ({\rm{isotropic}}) 
\label{eqoverview9}\\
&=&{1\over 2}m(\omega_{ox}^2 x^2+\omega_{oy}^2 y^2 +\omega_{oz}^2 z^2)\ 
({\rm anisotropic}).\label{eqoverview10}
\end{eqnarray}
Most current traps are either:
\begin{eqnarray}
{\rm{pancakes}} &,&  \omega_{oz} \gg \omega_{ox} (=\omega_{oy}) \nonumber\\
{\rm{cigars}} &,& \omega_{oz} \ll \omega_{ox} 
(=\omega_{oy})\label{eqoverview11}
\end{eqnarray}
and the trap frequencies are of the order $\omega_0 \sim 2\pi\times 
100$Hz.  In 1995, the first condensates were small $\sim 10^3$ atoms and 
$T_{BEC}\sim$ 100nK.  However in 1999, the condensates can be quite large 
$\sim 10^8$ atoms at $T_{BEC}\sim\mu$K.  These have a size $\sim$ many 
microns, which can be easily seen optically .  When the condensates are 
small, the trap is turned off and cloud allowed to expand, and then 
measured by optical methods.  The results are simple to analyze if gas is 
non-interacting.  However, more analysis is needed to include the effects 
of interactions during expansion.

Early reports on atomic condensates discussed the system as an {\it ideal} 
Bose gas.  It was soon realized that even in these very dilute gases, the 
interactions played a crucial role.  Indeed,  since the 1960's, it has 
been understood that an interacting Bose gas is quite different from an 
ideal Bose gas.  In particular, interactions stabilize (or ``lock'') the 
phase of the condensate and allow {\it coherent} properties to emerge.  
(We recall that a free Bose gas has a condensate but is not a superfluid).
Of course, interactions are also crucial for cooling.  After  hot atoms 
are removed by rf-induced spin-flips, it is important that remaining atoms 
can quickly re-thermalize through collisions.

However, it is useful to first consider an {\it ideal} Bose gas in a trap, 
to illustrate some characteristic features.  For atoms in an external 
potential, we have
\begin{equation}
N = \sum_i f_0(\epsilon_i),\label{eqoverview12}
\end{equation}
where the Bose distribution is
\begin{equation}
f_0(\epsilon_i) = {1\over e^{\beta(\epsilon_i-\mu)}-1}. 
\label{eqoverview13}
\end{equation}
In a harmonic trap, the energy levels are:
\begin{equation}
\epsilon_i=\epsilon_{{n_x}{n_y}{n_z}} = (n_x+n_y+n_z+{3\over 
2})\hbar\omega_0,\label{eqoverview14}
\end{equation}
with $n_x, n_y, n_z = 0, 1, 2, 3\dots$.
The condensate is described by the ground state single-particle 
wavefunction 
\begin{equation}
\phi_0({\bf r})\sim e^{-r^2/2a^2_{HO}},\label{eqoverview15}\end{equation}
where the S.H. oscillator length is $a_{HO}\equiv\left({\hbar \over 
m\omega_0}\right)^{1\over 2}
\sim 1 \mu m$ in current traps.  Clearly $a_{HO}$ gives the ``size'' of 
condensate $n_{c0}({\bf r})= |\Phi_0({\bf r})|^2$ in a trapped gas. For 
the non-condensate density ${\tilde n}_0({\bf r})$ (often called the 
``thermal cloud''), we can use the {\it semiclassical limit}, since the 
thermal energy $(k_B T)$ is much larger than the spacing between the S.H. 
energy levels $(\hbar \omega_0).$  Then we have
\begin{eqnarray}
{\tilde n}_0({\bf r})&\sim& e^{-V_{ex}({\bf r})/k_B T} \nonumber\\
&=& e^{-r^2/2R_T^2},\label{eqoverview16}
\end{eqnarray}
where
\begin{equation}
R_T = {\sqrt{k_BT\over m\omega^2_0}} = a_{HO} \left({kT\over 
\hbar\omega_0}\right)^{1\over 2} \gg a_{HO}.\label{eqoverview17}
\end{equation}
Thus we see that the size of the thermal cloud $R_T$ is much larger than 
the condensate.
The signature for condensate is this {\it sharp} high density peak at the 
centre of the trap, which suddenly starts to grow out of the broad thermal 
distribution at the predicted transition temperature $T_{BEC}$.  As 
$T\rightarrow 0$ (effectively $T\ \lapx \ 0.4 \ T_{BEC}),$ the thermal 
cloud steadily disappears as all atoms go into the ground state 
$\phi_0({\bf r})$ given by (\ref{eqoverview15}), which is the macroscopic 
wavefunction for a non-interacting trapped Bose gas.  The temperature of 
the gas is measured from the temperature dependence of the tail of the 
thermal distribution given by (\ref{eqoverview16}).  

It is easy to calculate the transition temperature for atoms in an 
harmonic trap.  Separating out the condensate contribution in 
(\ref{eqoverview15}), we have
\begin{equation}
N=N_c+\sum_{i\ne 0} f^0(\epsilon_i); \ N_c = \int d{\bf r}|\phi_0({\bf 
r})|^2.\label{eqoverview18}\end{equation}
Making a change of variable $\beta \hbar\omega_0 n_x \equiv {\bar n}_x$, 
and using the continuum approximation, we have
\begin{eqnarray}
N-N_c&\simeq& \left({k_BT\over\hbar\omega_0}\right)^3\int^\infty_0 d{\bar 
n}_x
\int^\infty_0 d{\bar n}_y \int^\infty_0 d{\bar n}_z {1\over e^{({\bar n}_x 
+{\bar n}_y +{\bar n}_z)}-1} \nonumber\\
&=&\zeta(3) \left({k_B T\over \hbar\omega_0}\right)^3.  
\label{eqoverview19}
\end{eqnarray}
We note that the chemical potential $\mu_0 = {3\over 2}\hbar\omega_0,$ but 
${\hbar\omega_0\over k_B T}\ll 1$ and hence the zero point energy has 
been  neglected in (\ref{eqoverview19}).  Since $N_c = 0$ at $T_{BEC}$, we 
have
\begin{equation}
k_BT_{BEC} \simeq 0.94 (N^{1\over 3})\hbar\omega_0, \ \mbox{with}\ 
{N_c(T)\over N} = \left[1-\left({T\over 
T_{BEC}}\right)^3\right].\label{eqoverview20}
\end{equation}
One can improve on these simple estimates for $T_{BEC}$ and $N_c(T)$, but 
the corrections are only a few percent at best \cite{Dalgiorpit}.  

Interactions make a dilute, weakly-interacting Bose condensed gas into a 
full non-trivial many body problem, even though the system is very dilute 
and the interactions are weak. 
In a dilute gas, we need only consider binary collisions.  The real 
interatomic potential $v({\bf r})$ has a hard core with a radius of a few 
Angstroms and a weak, long-range attractive tail. In a dilute, very cold 
gas, we can approximate $v({\bf r})$ using the $s$-wave scattering length 
approximation effectively replacing $v({\bf r})$ by a pseudopotential 
\cite{Dalibard}
\begin{equation}
v({\bf r}) \Rightarrow {4\pi h^2\over m}a\delta({\bf r})\equiv 
g\delta{(\bf r}).  \label{eqoverview21}
\end{equation}
We require $a\ll$ average distance between atoms, or $na^3\ll 1,$ which is 
very well satisfied in these gases.  For alkali atoms, $v({\bf r})$ almost 
has a bound state of two atoms.  This quasi-bound state is very sensitive 
to the long range part of the potential, and thus the value of the 
$s-$wave scattering length $a$ can be very large.  Current values for 
atoms used in BEC experiments are:
\begin{eqnarray}
^{87}{\rm Rb} &:& a = 58 \stackrel{\circ}{\rm A} \nonumber \\
^{23}{\rm Na} &:& a=28 \stackrel{\circ}{\rm A} \nonumber\\
^7{\rm Li}&:& a=-14 \stackrel{\circ}{\rm A} \label{eqoverview22}
\end{eqnarray}

One can adjust the energy of the quasi-bound state and, as a result, 
change the value of s-wave scattering length $a$ with a small magnetic 
field.  Near a so-called {\it Feshbach resonance}, one can even change the 
interaction sign, going from repulsive $(a>0)$ to attractive $(a<0).$  
There is a lot of current work \cite{Phystodayart} trying to exploit this 
ability to change the interaction strength and sign by ``turning a 
knob'',  perhaps even making $a\rightarrow 0$!

The key equation for the macroscopic wavefunction for a $T=0$ condensate 
was written down and discussed by Pitaevskii \cite{LPPit} and Gross 
\cite{EPGross} in 1961:
\begin{equation}
i\hbar {\partial\Phi({\bf r}, t)\over\partial t} = 
\left[-{\hbar^2\nabla^2\over 2m}+V_{ex}({\bf r})+gn_c({\bf r}, 
t)\right]\Phi({\bf r}, t),\label{eqoverview23}
\end{equation}
where $n_c = |\Phi|^2.$  This equation describes the condensate atoms 
moving in dynamic self-consistent Hartree field produced by the 
condensate, 
\begin{equation} V_H({\bf r}, t) = 
\int d{\bf r}^\prime v({\bf r}-{\bf r}^\prime)n_c({\bf r}^\prime, t) = 
gn_c({\bf r}, t).\label{eqoverview24}\end{equation}  
The non-linear GP equation (\ref{eqoverview23}) will be the subject of 
Section~\ref{sec:Dynamics}.  Hundreds of papers have been written on it in 
the last four years.  For $T \ \lapx \ 0.4 \ T_{BEC},$ it describes both 
the static properties and the dynamic fluctuations (linear and non-linear) 
very well, usually within a few percent \cite{Dalgiorpit, Ingstrwie}.

To complete this brief introduction, I mention several other research 
topics that make trapped Bose-condensed gases so exciting:

\begin{enumerate}
\item Alkali atoms have several different atomic hyperfine states.  Apart 
from Cs, the alkali atoms have a nuclear spin $I=3/2$ and an electron spin 
$S = 1/2$.  Thus the total spin operator ${\bf F} = {\bf I}+{\bf S}$ has 
values of 1 and 2, leading to 8 possible atomic states.  One usually works 
with one of these atomic hyperfine states - trapped in a magnetic well.   
However, purely optical traps (MIT) using the dipole force of a laser beam 
can be used to trap low energy atoms in different hyperfine states. Thus 
one can now also deal with gases with several different states, ie, a 
multicomponent Bose gas. Moreover, one can induce transitions between 
different hyperfine states.  One sees that spin is a new degree of freedom 
in such multicomponent Bose-condensed gases \cite{Dalgiorpit, Ingstrwie}.

\item Independent of the special ``coherent'' features of a Bose 
condensate, these trapped gases give us a source of high density, very 
cold atoms.  Lene Hau \cite{Phystodayjul} has used this high density to 
slow down the speed of light to that of slow car ($\sim$ 40 km/hour) using 
self-induced transparency.  One can also switch on an optical lattice 
(produced by intersecting laser beams) on a trapped Bose gas 
\cite{Pzoller}.  Turning off  the magnetic trap, the low energy atoms will 
occupy the potential minima of this periodic lattice. This could not be 
done with high energy atoms since the dipole-induced potentials of the 
optical lattice are very weak.  With ultra-cold trapped fermions, one may 
also be able to produce a Hubbard model, of the kind extensively studied 
in connection with the cuprate-oxide high temperature 
superconductors.\end{enumerate}

We agree with the opinion of Pitaevskii \cite{LPPita} that the discovery 
of BEC in alkali gases ``can be considered as one of the most beautiful 
results of experimental physics in our century''.  The next two sections 
will flesh out this qualitative overview with some theoretical 
calculations on the collective oscillations at $T\ll T_{BEC}$ (pure 
condensate) and at finite temperatures $T\sim T_{BEC}$ (mixture of 
condensate and non-condensate) of these strange quantum ``wisps of 
matter''.

\section{DYNAMICS OF THE PURE CONDENSATE} 
\label{sec:Dynamics}
The theory of interacting Bose-condensed fluids is most usefully discussed 
using quantum field operators.  This procedure was formalized by Beliaev 
(1957) and developed by Bogoliubov \cite{Bog2}, Gavoret and Nozi\`{e}res 
\cite{Gavnoz}, Martin and Hohenberg \cite{Hohmar}, and others in the 
1960's \cite{review}.  
We recall:
\begin{eqnarray}
& {\hat\psi}^+({\bf r})= \mbox{creates atom at} \ {\bf r} \nonumber \\
&{\hat\psi}({\bf r}) =\mbox{destroys atom at} \ {\bf 
r}.\label{eqdynamics25}
\end{eqnarray}
These fields satisfy the usual Bose commutation relations, such as
\begin{equation}
\left[{\hat\psi}({\bf r}), {\hat\psi}^+({\bf r}^\prime)\right] = 
\delta({\bf r} - {\bf r}^\prime). \label{eqdynamics26}
\end{equation}
All observables can be written in terms of these quantum field operators, 
such as the interaction energy
\begin{eqnarray}
{\hat V}_{ext} &=& {1\over 2}\int d{\bf r} \int d{\bf r}^\prime 
{\hat\psi}^+ ({\bf r}^\prime){\hat\psi}^+({\bf r}) v({\bf r}-{\bf 
r}^\prime){\hat\psi}({\bf r}^\prime){\hat\psi}({\bf r})\nonumber \\
&=& {1\over 2}g \int d{\bf r} {\hat\psi}^+({\bf r}){\hat\psi}^+({\bf 
r}){\hat\psi}({\bf r}){\hat\psi}({\bf r}).\label{eqdynamics27}
\end{eqnarray}

The crucial idea due to Bogoliubov (1947) and later generalized by Beliaev 
is to separate out the condensate part
\begin{equation}
{\hat\psi}({\bf r}) = \left<{\hat\psi}({\bf r})\right> +{\tilde\psi}({\bf 
r}), \label{eqdynamics28}
\end{equation}
where
\begin{equation}
\left<{\hat\psi}({\bf r}) \right> \equiv \Phi({\bf r}) = \mbox{Bose 
macroscopic wavefunction}.\label{eqdynamics29}
\end{equation}
This quantity plays the role of the order parameter for the superfluid 
phase transition:
\begin{eqnarray}
\Phi({\bf r}) &= 0 \ \ \ \ T > T_c \nonumber \\
&\ne 0 \ \ \ \ T < T_c. \label{eqdynamics30}
\end{eqnarray}
We note that $\Phi({\bf r}) \equiv\sqrt{N_c} e^{i\theta}$ is a 2-component 
order parameter.  Clearly, $\Phi({\bf r})$ is not simply related to the 
many-particle wavefunction $\Psi({\bf r}_1, {\bf r}_2, \dots {\bf r}_N).$
The thermal average in $<{\hat\psi}({\bf r})>$ involves a small 
symmetry-breaking perturbation to allow $\Phi$ to be finite,
\begin{equation}
{\hat H}_{SB} = \int d{\bf r}\left[\eta ({\bf r}){\hat\psi}^+({\bf r}) + 
\eta^\ast({\bf r}) {\hat\psi}({\bf r})\right].\label{eqdynamics31}
\end{equation}

It is useful to make a few comments on the physics behind $\Phi({\bf r}, 
t)$.  $\Phi ({\bf r}, t)$ is a {\it coherent} state, with a ``clamped'' 
value of phase - rather than a Fock-state of fixed $N$, with no 
well-defined phase.  $\Phi({\bf r}, t)$ acts like a {\it classical} field, 
since quantum fluctuations are negligible when $N_c$ is large.  Probably 
P.W. Anderson deserves the greatest credit for understanding (in the 
period 1958-1963) the new physics behind working with a broken-symmetry 
state $\Phi({\bf r}, t)$, both in BCS superconductors and in superfluid 
$^4$He \cite{PWand}.   It captures the physics of the new phase of matter 
(such as the occurence of the Josephson effect) and the associated 
superfluidity.  The symmetry-breaking perturbation allows $<{\hat\psi}>$ 
to be finite.  More precisely, it allows the system to internally set up 
off-diagonal symmetry-breaking fields, which persist even when the 
external symmetry-breaking perturbation in (\ref{eqdynamics31}) is set to 
zero at the end $(\eta \rightarrow 0)$.  The same sort of physics is 
behind the BCS theory of superconductors.

The exact Heisenberg equation of motion for the field operator is
\begin{eqnarray}
i\hbar{\partial{\hat\psi}({\bf r}, t)\over\partial t} &=& 
\left[-{\hbar^2\nabla^2\over 2m}+V_{ex}({\bf r}) + \delta V({\bf r}, 
t)\right]
{\hat\psi}({\bf r}, t)\nonumber \\
&+&\eta({\bf r})+g{\hat\psi}^+ ({\bf r}, t){\hat\psi}({\bf r}, 
t){\hat\psi}({\bf r}, t),\label{eqdynamics32}
\end{eqnarray}
where $\delta V({\bf r}, t)$ is a small time-dependent driving potential.  
This gives an exact equation of motion for $\Phi({\bf r}, t) 
\equiv\left<{\hat\psi}({\bf r}, t)\right>,$
\begin{eqnarray}
i\hbar{\partial\Phi({\bf r}, t)\over\partial t} 
&=&\left[-{\hbar^2\nabla^2\over 2m}+V_{ex}({\bf r})
+\delta V({\bf r}, t)\right]\Phi({\bf r}, t)\nonumber \\
&+&\eta({\bf r})+g\left<{\hat\psi}^+ ({\bf r}, t){\hat\psi}({\bf r}, 
t){\hat\psi}({\bf r}, t)\right>,\label{eqdynamics33}
\end{eqnarray}
with
\begin{equation}
{\hat\psi}^+{\hat\psi}{\hat\psi} = |\Phi|^2\Phi + 
2|\Phi|^2{\tilde\psi}+\Phi^2{\tilde\psi}^+ + 
\Phi^\ast{\tilde\psi}{\tilde\psi}+2\Phi{\tilde\psi}^+{\tilde\psi} + 
{\tilde\psi}^+{\tilde\psi}{\tilde\psi}.\label{eqdynamics34}
\end{equation}
Taking the symmetry-breaking average, one finds
\begin{equation}
\left<{\hat\psi}^+{\hat\psi}{\hat\psi} \right> = n_c\Phi+{\tilde 
m}\Phi^\ast+2{\tilde n}\Phi+\left<{\tilde\psi}^+  
{\tilde\psi}{\tilde\psi}\right>,\label{eqdynamics35}
\end{equation}
where
\begin{eqnarray}
n_c ({\bf r}, t) &\equiv& |\Phi({\bf r}, t)|^2 = \mbox{condensate density} 
\nonumber\\
{\tilde n}({\bf r}, t) &\equiv&  \left<{\tilde\psi}^+({\bf r}, 
t){\tilde\psi}({\bf r}, t)
\right> = \mbox{non-condensate density} \nonumber\\
{\tilde m}({\bf r}, t) &\equiv&\left<{\tilde\psi}({\bf r}, 
t){\tilde\psi}({\bf r}, t)
\right> = \mbox{off-diagonal (anomalous) density}.\nonumber
\end{eqnarray}
Here we have separated out the {\it condensate} and {\it non-condensate} 
parts
\begin{equation}
{\hat\psi}\Rightarrow\left<{\hat\psi}\right>+{\tilde\psi}=\Phi 
+{\tilde\psi}.
\label{eqdynamics36}
\end{equation}
In general, the equation (\ref{eqdynamics33}) for $\Phi({\bf r}, t)$ is 
not closed - it is coupled to the dynamics of the non-condensate.  
However, in this Section we limit ourselves to $T\ll T_{BEC}$, where we 
can assume the non-condensate fraction is negligible, leaving 
\begin{equation}
i\hbar{\partial\Phi({\bf r}, t)\over\partial t} = 
\left[-{\hbar^2\nabla^2_r\over 2m}+V_{ex}({\bf r}) +g|\Phi({\bf r}, 
t)|^2\right]\Phi({\bf r}, t).\label{eqdynamics37}
\end{equation}
This is the famous time-dependent Gross-Pitaevskii equation for the 
condensate macroscopic wavefunction.  It gives a complete description of 
the dynamics of a coherent matter wave at $T=0.$

\subsection{Static condensate}
We first consider the time-dependent stationary GP equation, which has the 
solution
\begin{equation}
\left<{\tilde\psi}({\bf r}, t)\right>\equiv\Phi({\bf r}, t)=\Phi_0({\bf 
r})e^{-i\mu t/\hbar},
\label{eqdynamics38}
\end{equation}
where $\mu$ is the chemical potential.  The physics behind this can be 
seen from
\begin{eqnarray}
 \left<N-1|{\hat\psi}({\bf r}, t)|N\right>
&=& e^{iE_{N-1} t/\hbar} \left<N-1|{\hat\psi}({\bf r})|N\right>e^{-iE_N 
t/\hbar} \nonumber\\
&=&\left<N-1|\sqrt{N}|N-1\right>e^{-i(E_N-E_{N-1})t/\hbar}\nonumber\\
&=& \sqrt{N} e^{-i\mu t/\hbar}.
\label {eqdynamics39}\end{eqnarray}
Using (\ref{eqdynamics38}) in (\ref{eqdynamics37}) gives
\begin{equation}
i\hbar\left(-{i\mu\over\hbar}\right)\Phi_0({\bf 
r})=\left[-{\hbar^2\nabla^2\over 2m} +V_{ex}({\bf r})+g|\Phi_0({\bf 
r})|^2\right]\Phi_0({\bf r}).\label{eqdynamics40}
\end{equation}
The {\it static} GP equation for the static condensate wavefunction  
$\Phi_0({\bf r})$ is thus 
\begin{equation}
\left[-{\hbar^2\nabla^2\over 2m}-\mu+V_{ex}({\bf r}) +g|\Phi_0({\bf 
r})|^2\right]\Phi_0({\bf r})= 0.\label{eqdynamics41}
\end{equation}

A simple approximation in solving (\ref{eqdynamics41}) is to ignore the 
kinetic energy of the condensate, ie, neglect the $-{\hbar^2\nabla^2\over 
2m}$ term.  This is called the ``Thomas-Fermi''  approximation (TF) in the 
recent Bose gas literature \cite{Dalgiorpit}.  In this TF approximation,  
the static GP (\ref{eqdynamics41}) equation for $\Phi_0({\bf r})$ reduces 
to
\begin{equation}
\left[V_{ex}({\bf r}) + g|\Phi_0({\bf 
r})|^2\right]=\mu,\label{eqdynamics42}
\end{equation}
which is easly inverted to give the condensate density profile
\begin{eqnarray}
n_{c0}({\bf r}) &=& {1\over g}\left[\mu - V_{ex}({\bf r})\right] \nonumber 
\\
&=& {1\over g}\left[\mu - {1\over 2} 
m\omega^2_0r^2\right]>0.\label{eqdynamics43}
\end{eqnarray}
Clearly in the TF approximation, the ``size'' of the condensate is 
$R_{TF}$, where 
\begin{equation}
\mu = {1\over 2} m\omega^2_0 R^2_{TF}. \label{eqdynamics44}
\end{equation}
One finds $\mu$ from the condition $\int d{\bf r}n_c({\bf r})= N_c = N$, or
\begin{equation}
N_c = 4\pi\int^{R_{TF}}_0 dr r^2{1\over g}\left[\mu- {1\over 2} 
m\omega^2_0r^2\right].
\label{eqdynamics45}
\end{equation}
This gives
\begin{equation}
\mu = \hbar\omega_0 \left[15{Na\over a_{HO}}\right]^{2/5}; \ \ a_{HO} 
\equiv (\hbar/m\omega_0)^{1/2}. \label{eqdynamics46}
\end{equation}
We note that SH oscillator length $a_{HO}$ is the size of the ground state 
wavefunction (\ref{eqoverview15}) of an atom in a parabolic potential.  
Combining (\ref{eqdynamics45}) and (\ref{eqdynamics44}) gives
\begin{eqnarray}
R_{TF} &=& a_{HO}\left(15 {Na\over a_{HO}}\right)^{1/5} \nonumber \\
& \gg & a_{HO}, \  \mbox{if} \ {Na\over a_{HO}}\gg 1. \label{eqdynamics47}
\end{eqnarray}
We thus find the surprising result that interactions (while weak) spread 
out the ideal gas condensate $(R_{TF} \gg a_{HO})$ and decrease the 
density of the condensate at centre of trap.  The TF approximation for 
$n_{c0}({\bf r})$ is very good for large $N$, except for a small region 
near the edge of condensate $(\simeq R_{TF}).$  Experimental data confirms 
these GP predictions for the $n_{c0}({\bf r})$ condensate profile, 
emphasizing that the condensate is {\it not} simply the ground state 
wavefunction of the harmonic trap potential.  

\subsection{Dynamics of the condensate (collective modes)}
If we linearize around the static equilibrium value of the condensate
\begin{equation}
\Phi({\bf r}, t) = e^{-i\mu t/\hbar}\left[\Phi_0({\bf r}) +\delta\Phi({\bf 
r}, t)\right], \label{eqdynamics48}
\end{equation}
where $\delta\Phi\ll\Phi_0, $ we see that
\begin{equation}
i\hbar{\partial\Phi\over\partial t} = \left[-{\hbar\nabla^2\over 
2m}+V_{ex}({\bf r}) + g
\left[|\Phi_0|^2+\Phi^\ast_0\delta\Phi+\Phi_0\delta\Phi^\ast\right] 
\left[\Phi_0+\delta\Phi\right]e^{-{i\mu t\over 
\hbar}}\right]\label{eqdynamics49}
\end{equation}
which gives
\begin{equation}
i\hbar{\partial\delta\Phi({\bf r}, t)\over\partial t} = 
\left[-{\hbar\nabla^2\over 2m}+V_{ext}({\bf r}) + 2g
|\Phi_0|^2-\mu\right]\delta\Phi({\bf r}, t)+g\Phi^2_0\delta\Phi^\ast({\bf 
r}, t).\label{eqdynamics50}
\end{equation}
We also have a similar equation of motion for $\delta\Phi^\ast({\bf r}, 
t).$  Solving these two coupled equations with the ansatz
\begin{equation}
\delta\Phi({\bf r}, t) = u({\bf r})e^{-i\omega t} +v({\bf r}) e^{i\omega 
t},
\label{eqdynamics51}\end{equation}
we find two coupled ``Bogoliubov equations'' for the amplitudes $u$ and 
$v$ \cite{Fetwalecka, Edrupbur}:
\begin{eqnarray}
&\left[-{\hbar^2\nabla^2\over 2m}+V_{ex}({\bf r})-\mu+2g n_{c0} ({\bf 
r})\right]u({\bf r}) + gn_{c0} ({\bf r}) v({\bf r}) = E_i u({\bf r}) 
\nonumber \\
&\left[-{\hbar^2\nabla^2\over 2m}+V_{ex}({\bf r})-\mu+2g n_{c0} ({\bf 
r})\right]v({\bf r}) + gn_{c0} ({\bf r}) u({\bf r}) = -E_i v({\bf 
r})\label{eqdynamics52}
 \end{eqnarray}
Here $E_i\equiv \hbar\omega$ are the excitation energies of the 
condensate.  

The equations in (\ref{eqdynamics52}) have been solved numerically by 
several groups and the observed oscillations are in good agreement with 
these predictions \cite{Dalgiorpit, Edrupbur}. As an illustration of the 
physics, it is useful to solve (\ref{eqdynamics52}) for a uniform Bose 
gas.  In this case we have
\begin{eqnarray}
u({\bf r}) = ue^{i{\bf k}\cdot{\bf r}}\nonumber \\
v({\bf r}) = ve^{i{\bf k}\cdot{\bf r}},\label{eqdynamics53}
\end{eqnarray}
which gives
 \begin{eqnarray}
(\hbar\omega)^2 &=& \left[{\hbar^2k^2\over 2m}-\mu + 2g n_{c0}\right]^2 - 
\left[gn_{c0}\right]^2 \nonumber \\
&=& \epsilon^2_k + 2gn_{c0}\epsilon_k. \label{eqdynamics54}\end{eqnarray}
This is the famous Bogoliubov spectrum at $T=0$ \cite{Nnbog, Fetwalecka}.
Here we have used $\mu_0 = gn_{c0}$ discussed earlier (see 
(\ref{eqdynamics42})).  One finds
a phonon region at long wavelengths
\begin{equation}\hbar \omega_k = \hbar v_Bk \ ; \ v_B \equiv 
\left({gn_{c0}\over m}\right
)^{1/2}.\label{eqdynamics55}
\end{equation}
The cross-over from particle-like to this collective phonon region occurs 
at $k_c,$ where
\begin{equation}
{\hbar^2 k^2_c\over 2m}=2gn_{c0} \rightarrow k_c = \sqrt{4mn_{c0} g/\hbar}.
\label{eqdynamics56}\end{equation}
This shows how the interactions changes the qualitative nature of low 
energy exciations in a Bose-condensed gas.  This feature can be shown to 
stabilize superfluid motion against dissipation \cite{Gorkov}.

These oscillations of the condensate can be understood as excitations 
involving the {\it non-condensate}.  Using (\ref{eqdynamics36}), the 
Hamiltonian is given by \cite{AlFetter, Agriffin}
\begin{eqnarray}
{\hat H}-\mu{\hat N} &=& \int d{\bf r}{\tilde\psi}^+({\bf 
r})\left[-{\hbar^2\nabla^2\over 2m}+V_{ex}({\bf 
r})-\mu\right]{\tilde\psi}({\bf r}) \nonumber \\
&+&\int d{\bf r} \Phi^\ast_0 ({\bf r})\left[-{\hbar^2\nabla^2\over 2m} 
+V_{ex}({\bf r})-\mu\right]\Phi_0({\bf r})\nonumber \\
&+&2g\int d{\bf r}|\Phi_0({\bf r})|^2{\tilde\psi}^+({\bf 
r}){\tilde\psi}({\bf r})\nonumber\\
&+&{1\over 2}g\int d{\bf r}\Phi^2_0({\bf r}){\tilde\psi}^+({\bf 
r}){\hat\psi}^+({\bf r})\nonumber \\
&+&{1\over 2}g\int d{\bf r}\Phi^{\ast 2}_0({\bf r}){\tilde\psi}({\bf 
r}){\tilde\psi}({\bf r}).\label{eqdynamics57}
\end{eqnarray}
We can diagonalize the quadratic part of the Hamiltonian, using
\begin{equation}
{\tilde\psi}({\bf r}) = \sum_i\left[u_i({\bf r}){\hat\alpha}_i + 
v^\ast_i({\bf r})\alpha_i^+\right],\label{eqdynamics58}
\end{equation}
where $\left[{\hat\alpha}_i, {\hat\alpha}_j^+\right] = \delta_{ij}$ (Boson 
quasiparticles).  Thus one finds
\begin{equation}
{\hat H}-\mu{\hat N} = \mbox{const.} + \sum_i\hbar 
\omega_i{\hat\alpha}_i^+ {\hat\alpha}_i.\label{eqdynamics59}\end{equation}

This transformation shows how the non-condensate part of Hamiltonian can 
be reduced to a system of {\it non-interacting} quasiparticles with a 
spectrum identical to the condensate fluctuations.  This equivalence is 
easy to understand.  The condensate fluctuations
\begin{equation}
\delta\Phi \equiv \left<{\hat\psi}({\bf r})\right> - \Phi_0 
\label{eqdynamics60}\end{equation}
can be calculated to first order in the symmetry-breaking perturbation 
(\ref{eqdynamics31}), 
\begin{equation}
H_{sb} = \int d{\bf r} \left[\eta{\hat\psi}^+ + 
\eta^\ast{\hat\psi}\right].\label{eqdynamics61}\end{equation}
Then standard linear response theory \cite{Agrif} gives (schematically)
\begin{eqnarray}
\delta\Phi &\sim&  \int <[{\hat\psi}, H_{sb}]>\nonumber \\
&\sim& \int <[{\tilde\psi}, {\tilde\psi}^+]>\eta +<[{\tilde\psi}, 
{\tilde\psi}]>\eta^\ast.\label{eqdynamics62}\end{eqnarray}
This shows that the single-particle Green's functions of the 
non-condensate fields have the same spectrum as $\delta\Phi$.  This 
identity of the spectrum of density fluctuations and single-particle 
excitations is a characteristic signature of all Bose-condensed systems 
which persists at finite temperatures \cite{Agrif, Grif2}. 

One interesting collective oscillation is the dipole mode corresponding to 
{\it rigid} oscillation of the centre of mass of the static condensate 
profile, and predicted to have the trap frequency $\omega_0$.  This mode 
is described by
\begin{equation}
n_c({\bf r}, t) = n_{c0} ({\bf r} - \mbox{\boldmath$\eta$}(t)), \ \ \ 
{\dot{\mbox{\boldmath$\eta$}}}(t) = {\bf v}_c, 
\label{eqdynamics63}\end{equation}
where the time-dependent centre of mass satisfies
\begin{equation}{\partial^2\mbox{\boldmath$\eta$} (t)\over \partial t^2} = 
-\omega^2_0\mbox{\boldmath$\eta$}(t). \label{eqdynamics64}\end{equation}
This mode at frequency $\omega_0$ is special feature of a parabolic trap 
and is called the Kohn mode for in the case of interacting fermions 
\cite{Jfdobson}. This ``sloshing mode''  is used in BEC experiments to 
measure the natural frequency $\omega_0$ of the trap and it exists at 
finite temperatures as well (see Section \ref{sec:Coupled}).

\subsection{Quantum hydrodynamic formulation }
One often rewrites the time-dependent GP equation using the amplitude and 
phase variables \cite{Dalgiorpit}
\begin{equation}\Phi({\bf r}, t) = \sqrt{n_c} 
e^{i\theta}.\label{eqdynamics65}
\end{equation}
Inserting this into the GP equation (\ref{eqdynamics37}) and separating 
out the real and imaginary parts of the equation gives:
\begin{eqnarray}&& {\partial n_c({\bf r}, t)\over\partial t} + 
\mbox{\boldmath$\nabla$}\cdot n_c({\bf r}, t){\bf v}_c({\bf r}, t)= 0 , \  
\mbox{continuity equation} \label{eqdynamics66}\\
&&\hbar{\partial \theta({\bf r}, t)\over\partial t} = -\left[\mu_c({\bf 
r}, t)+{1\over 2}mv^2_c({\bf r}, t)\right], \mbox{Josephson equation} 
\label{eqdynamics67}\end{eqnarray}
Here the gradient of the phase is related to the superfluid velocity by
\begin{equation}m{\bf v}_c ({\bf r}, t) 
\equiv\hbar\mbox{\boldmath$\nabla$}\theta ({\bf r}, t)
\label{eqdynamics68}\end{equation}
and the condensate chemical potential is
\begin{equation}
\mu_c({\bf r}, t)\equiv-{\hbar^2\nabla^2\sqrt{n_c}\over 2m\sqrt{n_c}} + 
V_{ex}({\bf r}) + gn_c({\bf r}, t).\label{eqdynamics69}\end{equation}
Taking the gradient of (\ref{eqdynamics67}) gives
\begin{equation}
m\left({\partial{\bf v}_c\over\partial t} +{1\over 2}  
\mbox{\boldmath$\nabla$}{\bf v}_c^2\right) = - 
\mbox{\boldmath$\nabla$}\mu_c.\label{eqdynamics70}\end{equation}
The equations in (\ref{eqdynamics66}) and (\ref{eqdynamics70}) ``look'' 
like those in classical hydrodynamic theories. They show that the 
condensate can be described in terms of coherent motions involving two 
variables:
\begin{equation}
n_c ({\bf r}, t), \ {\bf v}_c({\bf r}, t)\label{eqdynamics71}\end{equation}
The Landau 2-fluid equations \cite{Land, Khalatnikov} reduce to these same 
equations at $T=0$  (where $\rho_s=\rho,\ \rho_n = 0),$ namely
\begin{eqnarray}
&{}&{\partial n\over\partial t} + \mbox{\boldmath$\nabla$}\cdot n{\bf v}_c 
= 0 \nonumber \\
&{}&m\left({\partial{\bf v}_c\over\partial t}+{1\over 2}\nabla{\bf 
v}_c^2\right) = 
-\mbox{\boldmath$\nabla$}\mu.\label{eqdynamics72}\end{eqnarray}
We will find these equations useful in Section~\ref{sec:Coupled}, where 
they complement the hydrodynamic equations describing the {\it 
non-condensate} in the collision-dominated region.  

This approach also allows a simple theory  developed by Stringari 
\cite{Sstringari} for linearized collective modes when we use the TF 
approximation.  Taking the time-derivative of (\ref{eqdynamics66}) gives 
$({\bf v}_{c0} = 0)$
\begin{equation}
{\partial^2\delta n_c\over \partial t^2} = - \mbox{\boldmath$\nabla$}\cdot 
\left[n_{c0} \left(
{\partial\delta{\bf v}_c\over \partial 
t}\right)\right].\label{eqdynamics73}
\end{equation}
Using (\ref{eqdynamics70}), we have
\begin{eqnarray} {\partial\delta{\bf v}_c\over \partial t} &=& -{1\over m}
 \mbox{\boldmath$\nabla$}\left[V_{ex}({\bf r}) + gn_{c0}({\bf r}) + 
g\delta n_c({\bf r}, t)\right]\nonumber \\
&=& -{g\over m} \mbox{\boldmath$\nabla$}\delta n_c ({\bf r}, t). 
\label{eqdynamics74}\end{eqnarray}
Combining this last result with (\ref{eqdynamics73}), we obtain the very 
useful Stringari equation of motion \cite{Sstringari}
\begin{equation}
{\partial^2\delta n_c\over \partial t^2} = 
\mbox{\boldmath$\nabla$}\cdot\left\{\left[\mu-{1\over 2}m\omega^2_0 
r^2\right]\mbox{\boldmath$\nabla$}\delta 
n_c\right\}.\label{eqdynamics75}\end{equation}
This describes the collective oscillations of the condensate in terms of a 
single differential equation.  As one example, the breathing mode of the 
condensate has a frequency $\hbar\omega = \sqrt{5}\hbar \omega_0.$  This 
example points out that in the TF limit (large $N_c$), the frequencies are 
independent of the interaction strength and the size of the condensate 
$N_c$.

We also note that using (\ref{eqdynamics74}), (\ref{eqdynamics75}) can be 
equally well rewritten in terms of the superfluid velocity ${\bf v}_c 
({\bf r}, t)$ defined in (\ref{eqdynamics68}).  This emphasizes that the 
condensate fluctuations are directly related to the existence of phase 
fluctuations.  Their existence may thus be viewed as ``evidence'' of 
superfluidity, the latter being always a consequence of the phase 
coherence of the macroscopic wavefunction given by (\ref{eqdynamics65}) 
\cite{Gorkov}.

The great thing about the collective oscillations of a condensate in a 
trapped gas is you can ``see'' them. A beautiful example from MIT is shown 
in Fig. 2 of Ref. \cite{Dalgiorpit}.  As Ketterle has remarked, these 
condensates are {\it robust} - one can kick them, shake them and these 
``wisps'' of Bose-condensed matter keep their integrity.

\subsection{Interference of coherent matter waves}
In the pioneering matter wave interference experiments done at MIT using a 
de-tuned cigar-shaped trap \cite{Wketterle}, one first destroys the 
condensate at the centre using laser beam.  Then the confining trap is 
turned off and the two condensates are allowed to expand and interfere.  
One observes nice interference fringes at the mid-point, as expected.
Using \begin{eqnarray} \Phi_{\mbox{system}}  &=& \Phi_A({\bf r}, t) + 
\Phi_B({\bf r}, t)\nonumber \\ 
&=&\sqrt{N_A}e^{{i\theta}_A}+\sqrt{N_B}e^{{i\theta}_B},\label{eqdynamics76}
\end{eqnarray}
the density is given by
\begin{eqnarray} n({\bf r}, t) &=& |\Phi_{\mbox{system}}|^2 \nonumber \\
&=& N_A + N_B + 2\sqrt{N_AN_B}\cos 
\Delta\theta,\label{eqdynamics77}\end{eqnarray}
where $\Delta\theta=\theta_A -\theta_B$.  In the region of interference, 
the density is low and hence interaction effects are {\it small} (ie, the 
$gn_c$ term is small in the GP equation).  Asymptotically, the solution of 
GP equation gives $\theta({\bf r}, t) = {mr^2\over 2\hbar t}.$  This 
implies \cite{Dalgiorpit}
\begin{equation}
\Delta\theta = {m\left(z+{d\over 2}\right)^2\over 2\hbar t} - 
{m\left(z-{d\over 2}\right)^2\over 2\hbar t} = {mzd\over\hbar 
t}\equiv{2\pi z\over\lambda(t)},\label{eqdynamics78}\end{equation}
where $\lambda(t)=2\pi\hbar t/md.$ This wavelength is in good agreement 
with experimental observations (See Fig. 2 of Ref. \cite{Wketterle}). 

A condensate described by $\Phi({\bf r}, t)$ may be viewed as a 
``classical'' matter wave, as recently emphasized by Pitaevskii and 
Stringari \cite{Lpitsstring}.  This is quite {\it different} from ordinary 
quantum deBroglie waves, since one can ignore quantum fluctuations (large 
$N_c$).  It is also quite different than ordinary (classical) macroscopic 
objects and electromagnetic waves, since $\Phi({\bf r}, t)$ is described 
by the GP equation (\ref{eqdynamics37}) which involves Planck's constant 
$\hbar$.  
Thus $\Phi({\bf r}, t)$ is a classical object which is described by a 
quantum equation!!  This promises to be a challenge for the quantum theory 
of measurement.

One can describe a two-component Bose gas (see Section~\ref{sec:Overview}) 
using coupled equations:
\begin{eqnarray}
i\hbar{\partial\Phi_1\over\partial t}&=& \left[-{\hbar^2\nabla^2\over 2m} 
+V_1({\bf r})+g_1|\Phi_1|^2+g_{12}|\Phi_2|^2\right]\Phi_1 \nonumber \\
  i\hbar{\partial\Phi_2\over\partial t}&=& \left[-{\hbar^2\nabla^2\over 
2m} +V_2({\bf r})+g_2|\Phi_2|^2+g_{12}|\Phi_1|^2\right]\Phi_2 
\label{eqdynamics79}.\end{eqnarray}
There are two {\it coupled} GP equations for {\it two} macroscopic wave 
functions.  Extensive studies \cite{Eacornhall} have been made at JILA 
using the two atomic hyperfine states of $^{87}$Rb: 
\begin{equation}
|F=1, m_F = -1 >, \ \ |F = 2, m_F = 1>. \label{eqdynamics80}
\end{equation}
In particular, one can study interesting interference effects between 
these coupled wavefunctions.  Recent work at JILA has used such 
two-component Bose fluids to produce the long sought-for vortex state in 
one of the components \cite{Matand}.

\section{COUPLED DYNAMICS OF THE CONDENSATE AND NON-CONDENSATE}
\label{sec:Coupled}

In this Section, we switch our attention from $T=0$ (ie, $T\ \lapx\ 0.4\ 
T_{BEC})$ to finite temperatures, where $N_c$ and ${\tilde N}$ are 
comparable in size.  We first consider how the GP equation of motion for 
$\Phi({\bf r}, t)$ is modified.  As a first step, we could use 
\cite{Agrif, Ezarnikgrif},
\begin{equation}
i\hbar{\partial\Phi({\bf r}, t)\over\partial t}= 
\left[-{\hbar^2\nabla^2\over 2m} +V_{ex}({\bf r})+gn_c({\bf r}, t)+ 
2g{\tilde n}({\bf r}, t)\right]\Phi({\bf r}, 
t).\label{eqcoupled81}\end{equation}
The last term takes into account that the condensate moves in the dynamic 
Hartree-Fock (HF) field produced by non-condensate atoms.  Immediately, 
one sees this generalized GP equation is no longer closed.  It requires a 
theory of the non-condensate fluctuations, as described by ${\tilde 
n}({\bf r}, t)= {\tilde n}_0({\bf r})+\delta{\tilde n}({\bf r}, t).$

A simpler version of (\ref{eqcoupled81}) is to treat the effect of the 
non-condensate as a {\it static} HF field \cite{Agriffin, Hutzargrif}:
\begin{equation}
2g{\tilde n}({\bf r}, t) \simeq 2g{\tilde n}_0({\bf r})
.\label{eqcoupled82}\end{equation}

\begin{enumerate}
\item This corresponds to treating the condensate moving in a static HF 
field of the non-condensate.
\item ${\tilde n}_0({\bf r})$ can be calculated (self-consistently) using 
the fluctuations of $\Phi$, as discussed in Section \ref{sec:Dynamics}. 
One finds that the depletion of the condensate at $T=0$ is only a few 
percent. 
\item This procedure gives reasonable results for the thermodynamic 
properties at finite temperatures, as discussed in the recent literature 
\cite{Dalgiorpit}.
Within the Thomas-Fermi approximation (good for $N\ \gapx\ 10^4$ atoms), 
the linearized version of (\ref{eqcoupled81}) using (\ref{eqcoupled82}) 
leads to the same Stringari equation at finite $T$ as the $T=0$ result in 
(\ref{eqdynamics75}).  Since the solutions of (\ref{eqdynamics75}) do not 
depend on the magnitude of the condensate, one concludes that the 
collective modes of the condensate will show no temperature dependence, 
even though the condensate is being thermally depleted.  This prediction 
does not appear to agree with experimental results when $T \ \gapx \ 0.6\  
T_{BEC}.$  This suggests that the {\it dynamics} of the non-condensate has 
to be included.
\end{enumerate}

We now go on to determining ${\tilde n}({\bf r}, t)$ directly by deriving 
a {\it quantum} Boltzmann equation for the single-particle distribution 
function of excited atoms $ f({\bf p}, {\bf r}, t)$ and then use:
\begin{equation}{\tilde n}({\bf r}, t)\equiv\int{d{\bf p}\over (2\pi)^3} 
f({\bf p}, {\bf r}, t).\label{eqcoupled83}\end{equation}
This procedure generalizes the approach of Boltzmann (1880's) for a 
classical gas, including the effect of binary collisions.  Clearly one 
must make some approximations!  One wants, initially, to find a {\it 
useful} kinetic equation that builds in {\it just} enough physics.  Here I 
will discuss such a quantum Boltzmann equation for a trapped 
Bose-condensed gas at finite temperatures, which has been extensively 
discussed by Zaremba, Nikuni and the author \cite{Zarnikgrif, 
Tnikzargrif}.  It is only valid in the so-called semi-classical limit, 
where it is sufficient to work with $f({\bf p}, {\bf r}, t)$.  The 
conditions are
\begin{equation}
k_BT\gg gn\ , \ k_BT\gg \hbar\omega_0 . \label{eqcoupled84}\end{equation}
In this domain, one also can assume that the important thermal excitaitons 
can be approximated by simple Hartree-Fock particle-like spectrum:
\begin{equation}
{\tilde\varepsilon}({\bf r}, t) = {p^2\over 2m}+ 2g \left[n_c({\bf r}, 
t)+{\tilde n}({\bf r}, t)\right] + V_{ex}({\bf r})\equiv {p^2\over 2m} + 
U({\bf r}, t).\label{eqcoupled85}\end{equation}
Clearly the resulting kinetic equation is not valid at very low 
temperatures, where the thermal excitations are described by a 
Bogoliubov-type spectrum.

We simply write down our quantum kinetic equation \cite{Zarnikgrif},
\begin{eqnarray}
{\partial f({\bf p}, {\bf r}, t)\over\partial t} &+&{{\bf p}\over 
m}\cdot\mbox{\boldmath$\nabla$}_r  
f({\bf p}, {\bf r}, t) - \mbox{\boldmath$\nabla$}_r 
U({\bf r}, t) \cdot\mbox{\boldmath$\nabla$}_p f({\bf p}, {\bf r}, t) 
\nonumber \\
&=& C_{22}[f] + C_{12}[f].\label{eqcoupled86}\end{eqnarray}
The right hand side describes how binary collisions effect the value of 
the single-particle distribution function $ f({\bf p}, {\bf r}, t)$.  
The effect of collisions between excited atoms in the non-condensate is 
described by:
\begin{eqnarray}
C_{22}[f] &=& {2g^2\over (2\pi)^5\hbar^7}\int d{\bf p}_2\int d{\bf 
p}_3\int d{\bf p}_4\delta({\bf p}+{\bf p}_2 - {\bf p}_3 - {\bf 
p}_4)\nonumber \\
&\times&\delta\left({\tilde\varepsilon}_p+{\tilde\varepsilon}_{p_2}-{\tilde\varepsilon}_{p_3}-{\tilde\varepsilon}_{p_4}\right)\nonumber 
\\
&\times&\left[(1+f)(1+f_2)f_3f_4 - 
ff_2(1+f_3)(1+f_4)\right].\label{eqcoupled87}\end{eqnarray}
This collision integral was discussed in detail in 1933 by Uehling and 
Uhlenbeck for $T>T_{BEC}$ \cite{Uehlbeck}.  
We recall that creating a Boson gives a factor $(1 + f)$ and destroying a 
Boson gives $f.$  In the classical high temperature limit, $f\ll 1$ and 
the collision integral $C_{22}$ considerably simplifies.

Where does $f({\bf p}, {\bf r}, t)$ come from in a microscopic derivation 
of (\ref{eqcoupled86})?  Basically we calculate the non-equilibrium 
real-time single-particle Green's functions of the non-condensate field 
operators (using the Kadanoff-Baym formalism \cite{Kadbaym}).  This gives 
(schematically)
\begin{equation}
g_1(1, 
1^\prime)\sim\left<{\tilde\psi}^+(1){\tilde\psi}(1^\prime)\right>,\label{eqcoupled88}\end{equation}

where $1\equiv {\bf r}_1, t_1 ; \  1^\prime={\bf r}^\prime_1, 
t^\prime_1.$  We then express this as $g_1({\bf r}, t; {\bf R}, T)$, where 
the {\it relative} and {\it centre of mass} coordinates are
\begin{eqnarray}
&{}&{\bf r}={\bf r}_1-{\bf r}^\prime_1 ; \ {\bf R} = {1 \over 2} 
\left({\bf r}_1+{\bf r}^\prime_1\right)\nonumber \\
&{}&t=t_1 - t_1^\prime \ \ \  T = {1\over 
2}\left(t_1+t_1^\prime\right).\label{eqcoupled89}\end{eqnarray}
Finally we Fourier transform
$g_1({\bf r}, t; {\bf R}, T)$ to find $g_1({\bf p}, \omega; R, T)$,
which gives the number of atoms at ${\bf R}, T$ with momentum ${\bf p}$ 
and energy $\hbar\omega.$  The single-particle Wigner distribution 
function is given by
\begin{equation}
f({\bf p}, {\bf R}, T) \equiv\int^\infty_{-\infty} d\omega g_1({\bf p}, 
\omega; {\bf R}, T).\label{eqcoupled90}\end{equation}
The Wigner distribution function  $f({\bf p}, {\bf R}, T)$ is the quantum 
generalization the classical single-particle distribution function 
\cite{Kadbaym}.  These remarks should indicate how we can go from an 
equation of motion for the single-particle Green's function (within a 
given self-energy approximation) to a kinetic equation for $f({\bf p}, 
{\bf R}, T)$.  We refer to the classic account given (for 
non-Bose-condensed gases) in the book by Kadanoff and Baym \cite{Kadbaym} 
for further details.  This powerful approach was generalized to uniform 
Bose-condensed gases by Kane and Kadanoff \cite{Kanekad}, and has been 
extended to trapped gases in recent work \cite{Tomgrif}.

In addition to $C_{22}$ collisions, we also have collisions which involve 
{\it one} condensate atom:
\begin{eqnarray}
C_{12}[f] &=& {2g^2\over (2\pi)^2\hbar^4}\int d{\bf p}_1 \int d{\bf p}_2 
\int d{\bf p}_3\delta (m{\bf v}_c +{\bf p}_1-{\bf p}_2-{\bf p}_3) 
\nonumber\\
&\times& \delta\left(\varepsilon_c 
+{\tilde\varepsilon}_{p1}-{\tilde\varepsilon}_{p2}-{\tilde\varepsilon}_{p3}\right)\times\left[\delta\left({\bf 
p}-{\bf p}_1)-\delta({\bf p}-{\bf p}_2)-\delta({\bf p}-{\bf 
p}_3\right)\right]\nonumber \\
&\times&\left[n_c(1+f_1)f_2f_3 
-n_cf_1(1+f_2)(1+f_3)\right].\label{eqcoupled91}\end{eqnarray}
Here the condensate atom has
\begin{eqnarray}
&{}&\mbox{energy:} \  \varepsilon_c = \mu_c + {1\over 2} mv_c^2 \ ; \ 
\mu_c = V_{ex} + gn_c +2g{\tilde n}\nonumber\\
&{}&\mbox{momentum:} \  {\bf p}_c = m{\bf v}_c 
\label{eqcoupled92}\end{eqnarray}
We note the {\it key} difference between $C_{12}$ and $C_{22}$ collisions:
\begin{itemize}
\item $C_{22}$ and $C_{12}$ conserve energy and momentum in collisions.
\item $C_{12}$ does {\it not} (but $C_{22}$ does) conserve the number of 
condensate atoms. $C_{12}$ describes how atoms are ``kicked'' in and out 
of condensate.
\end{itemize}

It turns out the generalized GP equation (\ref{eqcoupled81}) is also 
modified by a term related to $C_{12}[f]$.  This makes sense, since the 
$C_{12}$ collisions  modify the condensate wavefunction $\Phi({\bf r}, 
t).$  One finds the new GP equation is given by (see also 
Ref.\cite{Hcstoof}):
\begin{equation}
i\hbar{\partial\Phi({\bf r}, t)\over\partial t} = 
\left[-{\hbar^2\nabla^2\over 2m}+V_{ex} ({\bf r}) + gn_c({\bf r}, t) + 
2g{\tilde n}({\bf r}, t) -iR({\bf r}, t)\right]\Phi({\bf r}, 
t),\label{eqcoupled93}
\end{equation}
where
\begin{equation}
R({\bf r}, t)\equiv\int {d{\bf p}\over(2\pi)^3} {C_{12}[f({\bf p}, {\bf 
r}, t)]\over 2n_c({\bf r}, t)}.\label{eqcoupled94}\end{equation}
\vskip 2pt
\noindent More precisely, the dissipative {\it iR} term in 
(\ref{eqcoupled93}) arises from a three field correlation function in the 
exact equation of motion [see (\ref{eqdynamics33}) and 
(\ref{eqdynamics35})], and is given by \cite{Zarnikgrif}
\begin{equation}
\int{d{\bf p}\over (2\pi)^3} C_{12}[f] = {2g\over\hbar}\sqrt{n_c}\ 
Im\left<{\tilde\psi}^+{\tilde\psi}{\tilde\psi}\right>.\label{eqcoupled95}
\end{equation}

We have to solve for $f({\bf p}, {\bf r}, t)$ and $\Phi({\bf r}, t),$ 
treating $C_{12}[f]$ very carefully. We see that there will be an exchange 
of atoms between the ${\tilde n}({\bf r}, t)$ and $n_c({\bf r}, t)$ 
components through the $C_{12}$ collisions.  We can use these coupled 
equations for a variety of problems.  In these lectures, we will consider 
the collective oscillations of the combined system composed of condensate 
and non-condensate.  
It is useful to introduce {\it two} regimes to describe collective modes 
in interacting systems  \cite{Agrif, chap1}:
\begin{enumerate}
\item[I.]Collisionless (produced by {\it mean fields})
\begin{displaymath}
\omega\tau_R\gg 1 \ \ \mbox{or}\ \ T\ll \tau_R \ 
\left(\omega\equiv{2\pi\over T}\right)
\end{displaymath}
\item[II.] Hydrodynamic (produced by {\it collisions})
\begin{displaymath}
\omega\tau_R\ll 1 \ \ \mbox{or}\ \ T\gg \tau_R,
\end{displaymath}
\end{enumerate}
where $\tau_R$ is some appropriate relaxation time.
What should we use for $\tau_R$?  For a classical gas, this is the 
collision time \cite{Khuang}
\begin{equation}
{1\over\tau_c}={\tilde n}\sigma {\bar v}\label{eqcoupled96}\end{equation}
where
\begin{eqnarray*}
&{}&\sigma= 8\pi a^2 \ \mbox{(for Bose particles)}; \ a = s\mbox{-wave 
scattering length}.\\
&{}&{\bar v} \simeq \ \mbox{average velocity of atoms}\ 
\sim\sqrt{k_BT\over m}.\\
&{}&{\tilde n}  =\  \mbox{density of excited atoms}.
\end{eqnarray*}
Even for a Bose-condensed gas, taking $\tau_R \sim\tau_c$ is a reasonable 
first estimate \cite{Tnikzargrif}.
To get into the interesting hydrodynamic region $(\omega\tau_R\ll 1)$, we 
need small $\tau_R $, ie, a large  density ${\tilde n}$ or a large 
collision cross-section  $\sigma$ (perhaps using a Feshbach resonance, as 
discussed in Section~\ref{sec:Overview}).

Let us look at the kinetic equation (\ref{eqcoupled86}), writing it in the 
schematic form:
\begin{equation}{\hat{\cal L}} f = C_{22}[f] 
+C_{12}[f].\label{eqcoupled97}\end{equation}
In the collisionless region, we need only solve ${\hat{\cal L}} f = 0.$
In contrast, in the hydrodynamic region, the collisions are so {\it 
strong} they produce {\it local} equilibrium \cite{Khuang}.  That is, they 
force $f$ to satisfy $C_{22}[f]=0.$ The unique solution ${\tilde f}$ of 
this equation is well-known to be given by
\begin{equation}
{\tilde f}({\bf p}, {\bf r}, t) = {1\over e^{\beta[{({\bf p}-m{\bf 
v}_n)^2\over 2m}+U({\bf r}, t) - {\tilde\mu}(r, 
t)]}-1},\label{eqcoupled98}\end{equation}
where ${\bf v}_n$ is the average local velocity and ${\tilde\mu}$ is the 
local chemical potential of the thermal atoms.
This local equilibrium Bose distribution involves the local variables 
$\beta, {\bf v}_n, {\tilde\mu}$ and $U$, all of which depend on $({\bf r}, 
t)$.

Why must ${\tilde f}$ have the form in (\ref{eqcoupled98})?  To satisfy 
$C_{22}[f_1] = 0,$ we must have [see (\ref{eqcoupled87})]
\begin{equation}
(1+f_1)(1+f_2)f_3f_4 - 
f_1f_2(1+f_3)(1+f_4)=0,\label{eqcoupled99}\end{equation}
and this requires that
$f$ be given by the Bose distribution.
We have used the fact that
\begin{equation}
f(x)\equiv{1\over e^x-1}  = -[f(-x)+1]\label{eqcoupled100}\end{equation}
and that
\begin{equation}
\left.\begin{array}{cc}
&{\bf p}_1 + {\bf p}_2 = {\bf p}_3 + {\bf p}_4 \nonumber \\
&{}\nonumber\\
& 
{\tilde\varepsilon}_{p_1}+{\tilde\varepsilon}_{p_2}={\tilde\varepsilon}_{p_3}+

{\tilde\varepsilon}_{p_4}\end{array}\right\}
\ \mbox{energy and momentum 
conservation},\label{eqcoupled101}\end{equation}
where ${\tilde\varepsilon}_p={p^2\over 2m} + U({\bf r}, t).$ As an aside,  
using a kinetic equation is the most physical way of deriving the 
equilibrium Bose distribution.  The standard approach in statistical 
mechanics texts based on calculating a partition function does not bring 
out the reason {\it why}
\begin{equation}
f_{B,F} = {1\over e^{\beta(\epsilon-\mu)}\mp 
1}.\label{eqcoupled102}\end{equation}

However, while the fact that ${\tilde f}$ is given by the local 
equilibrium Bose distribution in $(\ref{eqcoupled98})$ ensures that 
$C_{22}[{\tilde f}]=0, $ one finds that $C_{12}[{\tilde f}] \ne 0.$  More 
precisely, we find from (\ref{eqcoupled91})

\begin{eqnarray}
&{}&\left[(1+{\tilde f}_1){\tilde f}_2{\tilde f}_3 - {\tilde 
f}_1(1+{\tilde f_2})(1+{\tilde f_3})\right] \nonumber\\
&{}&\propto \left[e^{-\beta[{\tilde\mu}-\mu_c-{1\over 2}m({\bf v}_n-{\bf 
v}_c)^2]}-1\right](1+{\tilde f}_1){\tilde f}_2{\tilde 
f}_3.\label{eqcoupled103}\end{eqnarray}
The expression in the square bracket only vanishes if the condensate and 
non-condensate are in {\it diffusive} equilibrium, which requires that
\begin{equation}
{\tilde\mu}=\mu_c+{1\over 2}m({\bf v}_n-{\bf 
v}_c)^2.\label{eqcoupled104}\end{equation}
When we {\it perturb} the system, this may not be true, ie, the two 
components may be out of diffusive equilibrium.

We can now derive hydrodynamic equations for non-condensate by taking 
moments of Boltzmann equation, the standard procedure used in classical 
gases \cite{Khuang}. The first moment gives a continuity equation with a 
source term:
\begin{eqnarray}
\int d{\bf p}\left\{{\cal L}{\tilde f} = C_{12}[{\tilde 
f}]\right\}\rightarrow {\partial{\tilde n}\over\partial t} = 
-\mbox{\boldmath$\nabla$}\cdot({\tilde n}{\bf v}_n)+
\Gamma_{12}[{\tilde f}], \label{eqcoupled105}\end{eqnarray}
where
\begin{eqnarray}
{\tilde n} &\equiv& \int{d{\bf p}\over (2\pi)^3} {\tilde f}({\bf p}, {\bf 
r}, t)\nonumber \\
{\tilde n}{\bf v}_n &\equiv& \int{d{\bf p}\over (2\pi)^3} {{\bf p}\over 
m}{\tilde f}({\bf p}, {\bf r}, t)\nonumber\\
\Gamma_{12}[{\tilde f}]&\equiv&\int{d{\bf p}\over(2\pi)^3} C_{12}[{\tilde 
f}].\label{eqcoupled106}\end{eqnarray}
More explicitly, we find
\begin{eqnarray}
\Gamma_{12}[{\tilde f}]&=& {2g^2n_c\over(2\pi)^5\hbar^7} 
[e^{-\beta[{\tilde\mu}-\mu_c-{1\over 2}m({\bf v}_n-{\bf 
v}_c)^2]}-1]\nonumber\\
&\times&\int d{\bf p}_1\int d{\bf p}_2\int d{\bf p}_3 \delta(m{\bf v}_c 
+{\bf p}_1 - {\bf p}_2 - {\bf p}_3) \nonumber \\
&\times&\delta(\varepsilon_c+{\tilde\varepsilon}_1-{\tilde\varepsilon}_2-{\tilde\varepsilon}_3)(1+{\tilde 
f}_1){\tilde f}_2{\tilde f}_3 \nonumber \\
&\equiv& \left[e^{-\beta[{\tilde\mu}-\mu_c-{1\over 2}m({\bf v}_n-{\bf 
v}_c)^2-1]}-1\right]{n_c\over\tau_{12}}.\label{eqcoupled107}\end{eqnarray}
We note that $\tau_{12}$ is a collision time \cite{Tnikzargrif} which 
describes the $C_{12}$ collisions between the C and N.C. atoms.  Combining 
(\ref{eqcoupled105}) with the continuity equation which results from 
(\ref{eqcoupled93}), 
\begin{equation}
{\partial n_c\over\partial t} = -\mbox{\boldmath$\nabla$}\cdot(n_c{\bf 
v}_c)-\Gamma_{12}[{\tilde f}], \label{eqcoupled108}\end{equation}
we see that the source term $\Gamma_{12}$ cancels out to give
\begin{equation}{\partial(n_c+{\tilde n})\over\partial 
t}=-\mbox{\boldmath$\nabla$}\cdot(n_c{\bf v}_c+{\tilde n}{\bf 
v}_n).\label{eqcoupled109}\end{equation}
Thus our theory gives the exact continuity equation for the total local 
density $n=n_c+{\tilde n}.$

Similarly, one finds
\begin{eqnarray}
&&\int d{\bf p}{\bf p}\left\{{\hat{\cal L}}{\tilde f}\right.= 
\left.C_{12}[{\tilde f}]\right\}\rightarrow m{\tilde n}\left({\partial{\bf 
v}_n\over\partial t}+{1\over 2}\mbox{\boldmath$\nabla$}{\bf 
v}^2_n\right)\nonumber\\
&& \ \ \ \ \  = -\mbox{\boldmath$\nabla$}{\tilde P}({\bf r}, t)-{\tilde 
n}\nabla U({\bf r}, t) -m({\bf v}_n-{\bf v}_c)\Gamma_{12}[{\tilde f}], 
\label{eqcoupled110}\end{eqnarray}
where the kinetic pressure is given  by
\begin{equation}
{\tilde P}({\bf r}, t) = {m\over 3}\int {d{\bf p}\over (2\pi)^3} ({\bf 
p}-m{\bf v}_n)^2{\tilde f}({\bf p}, {\bf r}, 
t).\label{eqcoupled111}\end{equation}
The second moment gives
\begin{eqnarray}
&&\int  d{\bf p} p^2\left\{{\cal L}{\tilde f} = C_{12} [{\tilde 
f}]\right\}\rightarrow{\partial{\tilde P}\over\partial 
t}+\mbox{\boldmath$\nabla$}\cdot({\tilde P}{\bf v}_n)\nonumber \\
&&\  = -{2\over 3}{\tilde P}\mbox{\boldmath$\nabla$}\cdot{\bf v}_n+{2\over 
3}\left[\mu_c +{1\over 2}m({\bf v}_n-{\bf 
v}_c)^2-U\right]\Gamma_{12}[{\tilde f}].\label{eqcoupled112}\end{eqnarray}

The detailed derivation of these results is not important here 
\cite{Zarnikgrif}.  The main thing is that the hydrodynamic equations 
(\ref{eqcoupled105}), (\ref{eqcoupled110}) and (\ref{eqcoupled112}) can be 
shown to describe the non-condensate in terms of three new 
``coarse-grained'' variables:
\begin{displaymath}{\tilde n}({\bf r}, t), {\bf v}_n({\bf r}, t) \ 
\mbox{and}\ {\tilde P}({\bf r}, t).\end{displaymath}
These are coupled to the two additional variables which describe the 
condensate:
\begin{displaymath}
n_c({\bf r}, t), \ {\bf v}_c({\bf r}, t). \end{displaymath}
We note that the two condensate equations of motion given by 
(\ref{eqdynamics70}) and (\ref{eqcoupled108}) are always ``hydrodynamic'' 
in form.  In contrast, it is only in the collision-dominated region that 
the non-condensate dynamics can be described in terms of a few collective 
variables.  
We thus have 5 variables and 5 equations, which form a closed system. 
Both components exhibit {\it coupled}, {\it coherent} collective motions.  
This is the essence of two-fluid superfluid behaviour \cite{Land, 
Khalatnikov}, a new unexplored frontier in trapped Bose gases. 

What is {\it new} about the two-fluid hydrodynamic equations derived above 
is the role of the source term $\Gamma_{12}[{\tilde f}].$  In a linearized 
theory expanded around the {\it static} equilibrium Bose distribution 
${\tilde f}_0$
 (where $\Gamma_{12}[{\tilde f}_0]$ vanishes),  one finds 
\cite{Zarnikgrif, Tnikzargrif}
\begin{equation}
\Gamma_{12}[{\tilde f}]=\delta\Gamma_{12}[{\tilde f}]=-{\beta_0 
n_{0c}\over\tau^0_{12}}\delta\mu_{diff},\label{eqcoupled113}\end{equation}
where $\mu_{diff}\equiv{\tilde\mu}-\mu_c$.  We find an equation of motion 
of the kind
\begin{equation}
{\partial\delta\mu_{diff}\over\partial t} = 
-{\delta\mu_{diff}\over\tau_\mu}+\dots,\label{eqcoupled114}
\end{equation}
where (see Eq.(87) in Ref. \cite{Zarnikgrif}).

\begin{equation}
{1\over\tau_\mu}\equiv
\left({gn_{c0}\over 
k_BT}\right){1\over\sigma}{1\over\tau_{12}^0}.\label{eqcoupled115}\end{equation}

Here $\sigma$ involves various static equilibrium thermodynamic 
functions.  The new relaxation time $\tau_\mu$ (which we can calculate!) 
determines {\it how fast} ${\tilde\mu}\rightarrow\mu_c$, ie, how fast we 
reach diffusive equilibrium between the condensate and non-condensate.  We 
can have
\begin{equation}
\left.\begin{array}{cc}
&\omega\tau_{22}\ll 1\nonumber \\
& \omega\tau_{12}\ll 1\end{array}\right\}
\ \mbox{required for hydrodynamics}\label{eqcoupled116}\end{equation}
but simultaneously
\begin{equation}
\omega\tau_\mu\gg 1\label{eqcoupled117}\end{equation}
near $T_{BEC}$, where $n_{c0}\rightarrow 0$.  Our hydrodynamic equations 
predict the existence of a new relaxational mode \cite{Zarnikgrif, 
Tnikzargrif}
\begin{equation}
\omega\simeq -i/\tau_\mu.\label{eqcoupled118}\end{equation}
This mode is not included in the standard Landau 2-fluid equations (where 
$\rho_s$ and $\rho_n$ are assumed to be always in local equilibrium with 
each other).
\vskip 4pt

In a uniform gas, the two-fluid hydrodynamic equations give two normal 
mode solutions \cite{grifzar, Zarnikgrif}:
\begin{itemize}
\item First sound (oscillation of the non-condensate mainly)
\begin{equation}\omega = u_1k\ , \ u^2_1 \simeq{5\over 3}{{\tilde 
P}_0\over m{\tilde n}_0}\sim{kT\over m}.\label{eqcoupled119}\end{equation}
\item Second sound (oscillation of the condensate mainly)
\begin{equation}
\omega=u_2k\ , \ u^2_2 \simeq {gn_{c0}\over 
m}.\label{eqcoupled120}\end{equation}
\end{itemize}
We note the second sound mode is the hydrodynamic version of famous $T=0$ 
Bogoliubov phonon mode discussed in Section \ref{sec:Dynamics}.  It is the 
``soft mode'' at $T_{BEC}$.  This second sound mode couples to the new 
relaxational mode given in (\ref{eqcoupled118}) and is damped as a result, 
the maximum damping occuring when $\omega\tau_\mu=1.$

In  a trapped gas, we can work out the spectrum of hydrodynamic 
oscillations $(\sim e^{-i\omega t})$.  Both the condensate and 
non-condensate components have the same frequency.  The most interesting 
one is the dipole mode, described by
\begin{eqnarray}
&{}&{\tilde n}({\bf r}, t) = {\tilde n}_0 ({\bf 
r}-\mbox{\boldmath$\eta$}_n(t)), \ \ \ {\dot{\mbox{\boldmath$\eta$}}}_n(t) 
= {\bf v}_n \nonumber \\
&{}&n_c({\bf r}, t) = n_{c0} ({\bf r}-\mbox{\boldmath$\eta$}_c(t)), \ \ \ 
{\dot{\mbox{\boldmath$\eta$}}}_c(t) = {\bf v}_c. 
\label{eqcoupled121}\end{eqnarray}
One finds there are two modes of this kind \cite{Ezarnikgrif, Zarnikgrif}:
\begin{itemize}
\item In-phase (or Kohn) mode, where $\mbox{\boldmath$\eta$}_n= 
\mbox{\boldmath$\eta$}_c$ and $\omega=\omega_0$ (trap frequency). It is 
the finite temperature version of the sloshing mode described by 
(\ref{eqdynamics63}) and (\ref{eqdynamics64}).  We note that this mode is 
generic (occuring in both the hydrodynamic and collisionless limit) and is 
not damped \cite{Zarnikgrif}.
\item Out-of-phase dipole mode, with  $\mbox{\boldmath$\eta$}_n\ne 
\mbox{\boldmath$\eta$}_c$ and in {\it opposite} directions.  The frequency 
of this mode is different from the trap frequency.  This out-of-phase mode 
is of special interest since it is the analogue of the out-of-phase second 
sound mode in superfluid $^4$He. \end{itemize}

We conclude this Section with some remarks:
\begin{enumerate}\item The {\it specific calculation} sketched above is 
also of interest in the general field of non-equilibrium statistical 
physics.  It describes the detailed dynamics of a system with a 
two-component order parameter self-consistently coupled to a gas of 
excitations based on a fully microscopic theory.
\item More work is needed to extend our analysis to {\it low} but finite 
temperatures and also into the critical region {\it very close} to 
$T_{BEC}$.  In both cases, our simple Hartree-Fock particle-like thermal 
excitation spectrum (\ref{eqcoupled85}) is no longer valid.

\item The {\it classical}
  kinetic theory of gases has been a rich subject in mathematical physics 
in the twentieth century, with well-known contributions by people like 
Boltzmann, Hilbert, Enskog, Chapman, Uhlenbeck and Burnett.  These new 
equations of motion for a Bose-condensed gas promise to yield a lot of new 
physics in the next century - and surprises, as our work in this Section 
has already shown.
\end{enumerate} 

\section*{Acknowledgements}

The work in Section \ref{sec:Coupled} is part of an on-going collaboration 
with Eugene Zaremba (Queen's University) and Tetsuro Nukuni (Tokyo 
Institute of Technology).
I would like to thank Yvan Saint-Aubin for his strong desire that BEC be 
represented at this CRM Summer School and insisting that I could fit it 
into my post-sabbatical schedule.  My work is supported by a research 
grant from NSERC.  I would also like to thank JILA (University of 
Colorado) for support during the time when I prepared these lectures.

\end{document}